\documentclass[12pt]{article}
\usepackage{graphicx,amsmath,amssymb}

\newcommand\pubnumber{Article 34 in eConf C1304143}
\newcommand\pubdate{\today}

\def\rvk{Centre for Astrophysics and Cosmology, Science 
Institute, University of Iceland, Dunhagi 5, 107 Reykjav\'ik, ICELAND}
\def\cop{Dark Cosmology Centre, Niels Bohr Institute, 
University of Copenhagen, Juliane Maries Vej 30, 2100 Copenhagen \O, DENMARK}
\def\lei{Department of Physics and Astronomy, University 
of Leicester, University Road, Leicester, LE1 7RH, UK}

\def\Title#1{\begin{center} {\Large #1 } \end{center}}
\def\Author#1{\begin{center}{ \sc #1} \end{center}}
\def\Address#1{\begin{center}{ \it #1} \end{center}}

\newcommand\pubblock{\rightline{\begin{tabular}{l} \pubnumber\\
         \pubdate  \end{tabular}}}
\newenvironment{Abstract}{\begin{quotation}  }{\end{quotation}}
\newenvironment{Presented}{\begin{quotation} \begin{center} 
             PRESENTED AT\end{center}\bigskip 
      \begin{center}\begin{large}}{\end{large}\end{center} \end{quotation}}
\def\Acknowledgements{\bigskip  \bigskip \begin{center} \begin{large}
             \bf ACKNOWLEDGEMENTS \end{large}\end{center}}

\def\beq{\begin{equation}}
\def\eeq#1{\label{#1}\end{equation}}
\def\eeqn{\end{equation}}

\def\beqa{\begin{eqnarray}}
\def\eeqa#1{\label{#1}\end{eqnarray}}
\def\eeqan{\end{eqnarray}}

\let\bar=\overbar

\def\O{{\cal O}}

\def\Dslash{\not{\hbox{\kern-4pt $D$}}}
\def\dslash{\not{\hbox{\kern-2pt $\del$}}}

\def\msb{{\bar{\ssstyle M \kern -1pt S}}}

\begin{document}
\begin{titlepage}
\pubblock

\vfill
\Title{The Redshift Distribution of the TOUGH Survey}
\vfill
\Author{P. Jakobsson, R. Chapman}
\Address{\rvk}
\Author{J.~Hjorth, D.~Malesani, J.~P.~U.~Fynbo, B.~Milvang-Jensen, T.~Kr\"uhler}
\Address{\cop}
\Author{N.~R.~Tanvir}
\Address{\lei}
\vfill
\begin{Abstract}
We present the redshift results from a Very Large Telescope program aimed 
at optimizing the legacy value of the \emph{Swift} mission: to 
characterize a homogeneous, X-ray selected, sample of 69 GRB host galaxies. 
19 new redshifts have been secured, resulting in a 83\% (57/69) redshift 
completion, making the survey the most comprehensive in terms of redshift 
completeness of any sample to the full \emph{Swift} depth, available to date. 
We present the cumulative redshift distribution and derive a conservative,
yet small, associated uncertainty. We constrain the fraction of \emph{Swift} 
GRBs at high redshift to a maximum of 10\% (5\%) for $z > 6$ ($z > 7$). The 
mean redshift of the host sample is assessed to be $\langle z \rangle \gtrsim 
2.2$. Using this more complete sample, we confirm previous findings that the
GRB rate at high redshift ($z\gtrsim3$) appears to be in excess of predictions 
based on assumptions that it should follow conventional determinations of the 
star formation history of the universe, combined with an estimate of its 
likely metallicity dependence. This suggests that either star formation at 
high redshifts has been significantly underestimated, for example due to a 
dominant contribution from faint, undetected galaxies, or that GRB production 
is enhanced in the conditions of early star formation, beyond those usually 
ascribed to lower metallicity.
\end{Abstract}
\vfill
\begin{Presented}
"Huntsville in Nashville" Gamma-Ray Burst Symposium\\
Nashville, USA,  April 14--18, 2013
\end{Presented}
\vfill
\end{titlepage}
\def\thefootnote{\fnsymbol{footnote}}
\setcounter{footnote}{0}

\section{Introduction}

We have secured GRB host galaxy information for a homogeneous sample of 69 
\emph{Swift} GRBs with a large program at the Very Large Telescope (VLT) 
\cite{tough1}. The sample has been carefully selected and obeys strict and 
well-defined criteria. To optimize the survey, we focused on systems with 
the best observability, which also have the best available information. The 
main results of The Optically Unbiased GRB Host (TOUGH) survey is presented 
in \cite{tough2,tough3,tough4,tough5}, including fundamental properties of 
the hosts, Ly$\alpha$ emission and new redshifts. 
\par
Here we present the TOUGH campaign for missing redshifts via VLT/FORS 
\cite{tough3} and VLT/X-shooter \cite{tough5}. We attempted spectroscopic 
observations of most TOUGH host candidates with $R \lesssim 25.5$\,mag that 
did not have a reported reliable redshift. In a nutshell, our approach is
to reach as high a redshift completion as possible and not rely on
pseudo-redshifts, e.g. \cite{tan}. We believe that examining the entire
iceberg is significantly more successful than using the tip of it for
extrapolation.

\section{Redshift Measurements and Constraints}

\begin{figure}
\includegraphics[bb=60 24 926 634,clip,width=0.98\textwidth]{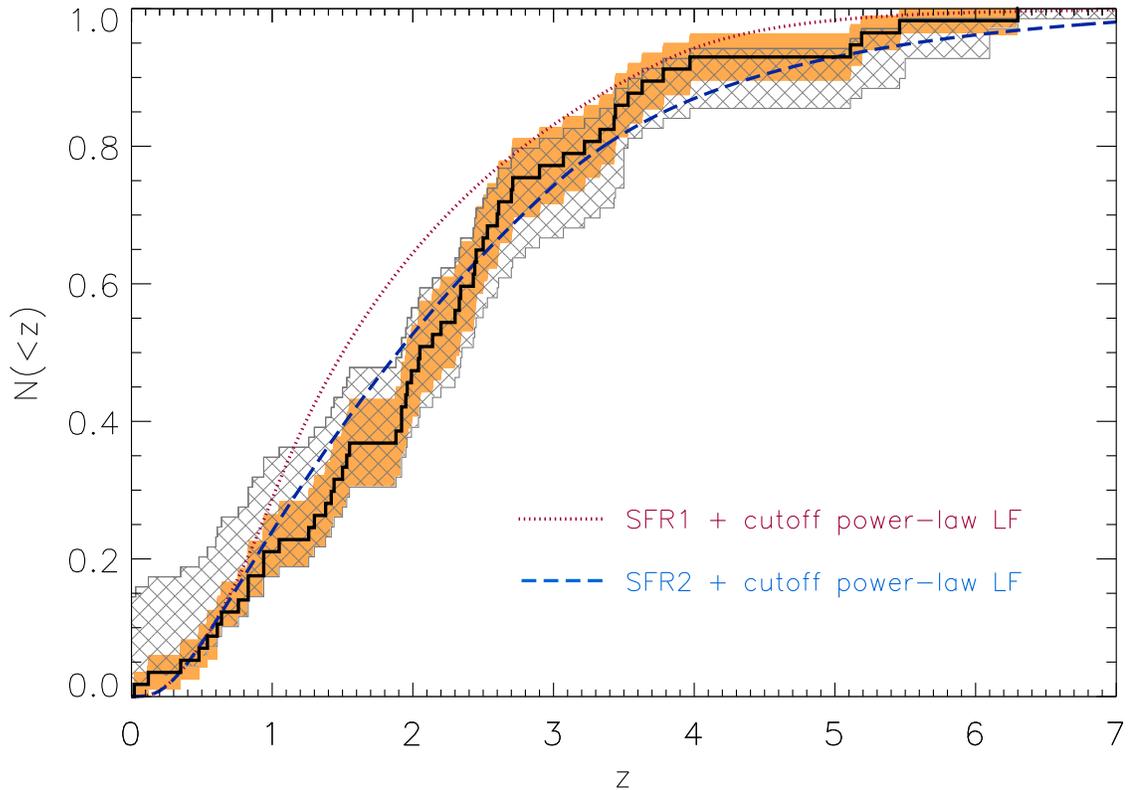}
\caption{{\bf Thick solid curve}: the cumulative fraction of GRBs as a
    function of redshift for the 57 \emph{Swift} bursts in the TOUGH
    sample with a measured redshift ($\langle z \rangle = 2.21$).
    {\bf Hatched region}: this is a conservative error region showing the
    systematic error on the thick solid curve.
    {\bf Shaded region}: statistical region showing the 1$\sigma$ sampling
    error band around the thick solid curve.
    {\bf Dotted curve}: the expected redshift distribution for \emph{Swift}
    observable long GRBs using the \emph{SFR1} history parameterization, i.e.
    the canonical SFR history discussed in \cite{hop} (see the main text).
    {\bf Dashed curve}: the same redshift distribution for the \emph{SFR2}
    history parameterization, i.e. a model where the SFR history remains 
    constant beyond $z \sim 3$ \cite{schmidt} (see the main text). 
    Updated from \cite{tough3}.}
    \label{zdist.fig}
\end{figure}

We have obtained 19 new host redshifts; Fig.\ref{zdist.fig} shows the 
cumulative redshift distribution of the 57 TOUGH bursts with a measured 
redshift. Also plotted is a conservative systematic error band (hatched 
region) containing information for all the 69 TOUGH bursts. The shaded region 
represents the likely statistical (1$\sigma$ standard error of the sample) 
uncertainty of the measured redshift distribution under the assumption that it 
is a true random sample of the overall population.  The sampling error and the
conservative systematic error region are shown separately to clearly 
illustrate that incompleteness dominates the sample, and more is gained by 
reducing the systematics rather than increasing the sample size. Using both 
error regions we can set a conservative limit on the maximum number of 
\emph{Swift} bursts at $z > 6$ ($z > 7$): 10\% (5\%).
\par
The average (median) redshift of the 57 TOUGH bursts is $\langle z \rangle
= 2.21$ ($\tilde{z}  = 2.06$), significantly lower than the early \emph{Swift} 
results indicated \cite{palli}. This difference may simply reflect the 
comparatively small samples analyzed in that paper, but could also be due to 
an increased success in measuring redshifts $z < 2$ using weaker absorption 
lines in afterglow spectra, and via host galaxies. Indeed, the average of the 
19 new redshifts is $\langle z \rangle = 1.83$. The mean redshift of the whole 
TOUGH sample could be as low as $\langle z \rangle \sim 1.9$ (upper boundary 
of the hatched region) although it is unlikely that the majority of bursts 
with unknown redshifts would be located at very small distances. In fact, it 
is more probable that $\langle z \rangle \gtrsim 2.2$ since we have only 
targeted the brightest galaxies in the sample ($R \lesssim 25.5$\,mag) for 
spectroscopic follow-up.


\section{Modelling}

Illustrative model fits are presented in \cite{tough3} and described in 
detail there. We assume that the GRB rate follows the star-formation rate 
(SFR) history, and consider two different SFR history parameterizations which 
we label as follows. \emph{SFR1} is an update \cite{li} of the SFR history 
models of \cite{hop} to include data from \cite{bouwens, reddy}, combined with 
a low-metallicity modification following the prescription of \cite{langer}. 
\emph{SFR2} is model A from \cite{schmidt} which represents a SFR history 
which remains constant beyond $z \sim 3$. It may, for example, be considered 
a more extreme low-metallicity correction to the cosmic SFR. Or it may 
represent a correction \cite{kistler,virgili} to the high-redshift SFR as 
estimated from flux-limited surveys by the integration of galaxy luminosity 
functions (LFs) thus obtained. This would be due to a large amount of hidden 
star formation in faint, low-mass, and high specific SFR galaxies of the type 
that GRBs tend to be associated with at lower redshift.
\par
Modeling is performed in the standard manner \cite{guetta} to produce 
$\log N$-$\log L$ number count distributions for various parameters of the LF, 
which are then fit by $\chi^2$ minimization to the observed $\log N$-$\log L$ 
distribution of all \emph{Swift} bursts with peak photon flux 
$>1{\rm\, cm^{-2}\,s^{-1}}$. We emphasize that the redshift distribution is 
not part of this fitting procedure, but is always purely a result. In 
Fig.~\ref{zdist.fig}, we plot the redshift distributions from our best fitting 
models in comparison to the TOUGH redshift data.
\par
At face value, these results seem to imply that GRBs follow a cosmic SFR
history that is significantly enhanced at high redshift compared to estimates
from flux-limited surveys. Given what is known about GRB hosts, it is entirely
feasible that GRBs trace star formation at high redshift that would be 
undetectable by other means. It is of course also possible that the simple 
low-metallicity enhanced SFR parameterization used in the \emph{SFR1} model is
inadequate, or that the LF could have a more complex form and/or evolve with 
redshift. It should also be noted that \cite{elliott} find that there is no 
strong preference for a metallicity cut.

\section{Discussion}

It is possible that star formation at high redshifts has been significantly 
underestimated. Even at $z \sim 2$ it appears that the galaxy LF has a 
substantially steeper faint-end slope than locally \cite{reddy09}, while 
recent LF studies in the Hubble Ultra-Deep Field have concluded that at 
$z \gtrsim 7$ so-far undetected galaxies are likely to completely dominate 
the total star formation activity \cite{bouwens11,nial}. Alternatively, it 
could be that GRB production is substantially enhanced in the conditions of 
early star formation, beyond the metallicity-dependent rate correction already 
applied. In the long run, large complete samples of GRB redshifts should shed 
light on whether the GRB rate is proportional to SFR or whether other effects 
play an important role.
\par
We have now reached a point in GRB research where a single burst rarely
elucidates and illuminates our general understanding of the field. It is
important to focus on well-defined samples and population studies, where
systematics and biases can be minimized. \emph{Swift} has made it possible to
build such a sample and thanks to new available instrumentation, such as the
VLT/X-shooter \cite{vernet}, we can continue to follow this track into the 
future.

\Acknowledgements
P.J. acknowledges support by a Project Grant from the Icelandic Research Fund.
The Dark Cosmology Centre is funded by the Danish National Research 
Foundation. This work has made use of the University of Hertfordshire Science 
and Technology Research Institute high-performance computing facility.

\end{document}